\documentclass[prl,aps,showpacs,twocolumn,superscriptaddress]{revtex4}
\usepackage{graphicx,bm}
\usepackage{amssymb}
\usepackage{epsfig}
\usepackage{epsf}
\usepackage[usenames]{color}

\begin{document}

\title{Magnonic black holes}

\author{A. Rold\'an-Molina}
\affiliation{
	Departamento de F\'isica, Facultad de Ciencias F\'isicas y Matem\'aticas, Universidad de Chile, Casilla 487-3, Santiago, Chile and Centro para el Desarrollo de la Nanociencia y la Nanotecnolog\'ia, CEDENNA, Avda. Ecuador 3493, Santiago 9170124, Chile}

\author{A.S. Nunez}
\affiliation{Departamento de F\'isica, Facultad de Ciencias F\'isicas y Matem\'aticas, Universidad de Chile, Casilla 487-3, Santiago, Chile}

\author{R.A. Duine}
\affiliation{Institute for Theoretical Physics, Utrecht
University, Leuvenlaan 4, 3584 CE Utrecht, The Netherlands}

\affiliation{Department of Applied Physics, Eindhoven University of Technology, P.O. Box 513, 5600 MB Eindhoven, The Netherlands}

\date{\today}

\begin{abstract}
We show that the interaction between spin-polarized current and magnetization dynamics can be used to implement black-hole and white-hole horizons for magnons - the quanta of oscillations in the magnetization direction in magnets. We consider three different systems: easy-plane ferromagnetic metals, isotropic antiferromagnetic metals, and easy-plane magnetic insulators. Based on available experimental data, we estimate that the Hawking temperature can be as large as $1$ K. We comment on the implications of magnonic horizons for spin-wave scattering and transport experiments, and for magnon entanglement. 
\end{abstract}


\maketitle

% definitions
\def\bx{{\bm x}}
\def\bk{{\bm k}}
\def\bK{{\bm K}}
\def\bq{{\bm q}}
\def\br{{\bm r}}
\def\bp{{\bm p}}
\def\bM{{\bm M}}
\def\bs{{\bm s}}
\def\bB{{\bm B}}
\def\bj{{\bm j}}
\def\bF{{\bm F}}
\def\id{{\rm d}}

\def\br{{\bm r}}
\def\bv{{\bm v}}

\def\half{\frac{1}{2}}
\def\args{(\bm, t)}

{\it Introduction} --- Hawking's 1974 prediction \cite{Hawking1974} that black holes evaporate by radiating particles with a thermal spectrum has triggered an enormous amount of scientific research and debate. It showed that black holes have a temperature - now called the Hawking temperature -  confirming earlier ideas by Bekenstein on black-hole entropy and black-hole thermodynamics \cite{Bekenstein1973}. Computing the black-hole entropy from a microscopic statistical-physics description has been a key test for candidates of quantum-gravity theories ever since \cite{strominger1996}. At the foundational level, the scaling of the black-hole entropy with area rather than volume led to the formulation of the so-called holographic principle \cite{holographic} and to debates concerning the black-hole information paradox \cite{informationparadox}. 

Despite these developments, Hawking radiation from gravitational black holes has not been observed yet. This is in part due to the low Hawking temperatures associated with astronomical black holes. Creating small black holes - which should have a higher Hawking temperatue - seems experimentally impossible, and, if one succeeded, they would evaporate rapidly. To circumvent such problems and to shed light on conceptual issues in the theoretical treatment of Hawking radiation, such as the so-called transplanckian problem, Unruh \cite{Unruh} suggested experimentally creating black-hole-horizon analogues. A black-hole horizon for sound waves in a flowing medium is created by a transition from subsonic to supersonic flow, such that waves along the flow and incoming from the subsonic region cannot escape from the supersonic region. A white-hole horizon is then a region where the flow changes from supersonic to subsonic. In this case, a wave travelling against the flow from the subsonic region cannot penetrate the supersonic part. Unruh's original proposal concerned waves in water. This system cannot be driven into the quantum regime where the temperature is much lower than the Hawking temperature. Nonetheless, by measuring the energy-dependence of reflection and transmission amplitudes of waves scattering off the horizon in the classical regime, the Hawking spectrum can be determined up to normalization as the underlying physics is linear. This was experimentally implemented in Ref.~\cite{Weinfurther2011}.

Unruh's work motivated theoretical proposals for black-hole-horizon analogues based on different systems in different regimes \cite{book0,book}. These include theoretical proposals for superfluid helium \cite{Jacobson1998}, atomic Bose-Einstein condensates \cite{garay2000}, light in dispersive media \cite{leonhardt2000}, electromagnetic waveguides \cite{Schutzhold2005}, ultracold fermions \cite{Giovanazzi2005}, trapped-ion rings \cite{horstmann2010}, exciton-polariton condensates \cite{Solnyshkov2011}, light in non-linear liquids \cite{elazar2012}, and, most recently, Weyl semi-metals \cite{volovik2016}. Experimental observations of various aspects of horizons have been observed in Bose-Einstein condensates \cite{Steinhauer}, optical systems \cite{optical}, and exciton-polariton condensates \cite{nguyen2015}. The essential ingredients for analogue horizons are linearly dispersing waves at long wavelengths and a background flow velocity which can exceed the velocity of the waves.

In this Letter, we propose a solid-state realization of a black-hole-horizon analogue. We outline how to use spin transfer torques, i.e, torques arising from the interplay between spin current and magnetization dynamics \cite{ralph2008}, to implement a black hole for magnons - the quanta of spin waves. In short, our proposal is based on the result that a spin-polarized electric current through a magnetic conductor interacts with the magnetization dynamics to give the spin waves a Doppler shift with effective ``spin-drift" velocity ${\bm v}_s$ - as was experimentally detected in Ref.~\cite{Vlaminck2008}. ``Supersonic" and ``subsonic" regions are then regions were the velocity ${\bm v}_s$ is larger or smaller (in absolute value) than the spin-wave velocity $c$. 

Our proposal is distinct from other implementations in that the background flow for the excitations is not provided by a moving medium but rather by a separate ``fluid"  - the spin current - that is controlled electrically and interacts with the magnetization and its excitations. In addition, the dissipation and, in particular, the dissipative coupling of spin current to the magnetization - or in the language of analogue gravity: the dissipative coupling between excitations and background flow - is well understood, which facilitates understanding its interplay with Hawking radiation. Moreover, the interaction between solitonic excitations - magnetic domain walls - and spin current allows for control over the position of domain walls as has been demonstrated experimentally with the long-term goal of building the magnetic ``race-track" memory \cite{parkin2015}. This allows for a controlled study of the interaction between domain walls and the magnonic horizon. 

From a practical point-of-view this system is attractive as it can be embedded in a device, can be electrically contacted, and has properties that are controlled by magnetic fields and electical currents. While this facilitates experiments, a magnonic black-hole horizon may in the longer term also serve as an on-chip resource of entangled magnons for magnon-based quantum computation and information purposes \cite{adrianov2014} - as pairs of Hawking particles emitted from the horizon are entangled \cite{Hawking1974,busch2014}.  This was experimentally demonstrated very recently in Bose-Einstein condensates \cite{Steinhauer}. Below we outline our proposals, provide estimates for their Hawking temperatures, and comment on experimental implications in the classical and quantum regime.  We discuss three systems: ferromagnetic metals, antiferromagnetic metals, and magnetic insulators. In the first of these, experiments have advanced furthest while experiments on antiferromagnetic metals and insulators are rapidly catching up.

{\it Easy-plane ferromagnetic metal} --- We consider a ferromagnetic metal far below its Curie temperature such that the unit vector ${\bm n} (\bx,t)$  along  the direction  of the magnetic order parameter  is the appropriate degree of freedom at low energies and long wavelengths. Spintronics research over the past decade \cite{slonczewski1996, berger1996,tsoi1998,myers1999,bazaliy1998,rossier2004,zhang2004,barnes2005,tserkovnyak2006,kohno2006,piechon2006,duine2007,ralph2008,tatara2004,tatara2008} has established that, in the presence of a steady-state transport current, it obeys the Landau-Lifshitz-Gilbert equation with spin transfer torques given by
\begin{equation}
\label{eq:LLGwithSTTs} \left( \frac{\partial }{\partial t} + {\bm
v}_{\rm s} \cdot \nabla  \right) {\bm n}  - {\bm n} \times
{\bm
 H}_{\rm eff} = - \alpha  {\bm n} \times  \left( \frac{\partial }{\partial
 t}+ \frac{\beta}{\alpha} {\bm
v}_{\rm s} \cdot \nabla \right) {\bm n}~,
\end{equation}
provided that spin-orbit coupling is not very strong. In this equation, the velocity ${\bm v}_s = -g P \mu_B {\bm j}/2eM_s$ that is proportional to the electrical transport-current density ${\bm j}$ parameterizes the reactive and dissipative spin transfer torques, corresponding to the terms proportional to ${\bm v}_s$ on the left and right-hand side of the above equation, respectively. Here, $g$ is the Land\'e factor, $P$ the spin polarization of the current, $\mu_B$ the Bohr magneton, $e$ minus the electric charge, and $M_s$ the saturation magnetization. The Gilbert damping parameter is given by $\alpha$. Usually, the dissipative coefficients $\beta \sim \alpha$ because of approximate Galileian invariance, and are of the order $10^{-2}$. The above equation accurately describes  experiments on current-driven domain wall motion  in permalloy and other magnetic materials \cite{grollier2003,tsoi2003,yamaguchi2004,klaui2005,beach2006,hayashi2007,yamanouchi2004}, and also predicts the spin-wave Doppler shift that was measured in Ref.~\cite{Vlaminck2008}. 

The effective field ${\bm H}_{\rm eff} = -\delta E/(\hbar \delta {\bm n})$ is determined as the functional derivative of the energy $E[{\bm n}]$ and acquires contributions from exchange, anisotropies and external fields. Here, we consider an easy-plane configuration and a field in the $z$-direction such that
\begin{equation}
\label{eq:energy}
  E[{\bm n}] = \int \frac{d {\bm x}}{a^3} \left[ -\frac{J_s}{2} {\bm n} \cdot \nabla^2 {\bm n} + \frac{K}{2} n_z^2 + B n_z \right]~,
\end{equation}
with $a^3$ the volume of a unit cell, $J_s$ the spin stiffness, and $B$ the external field (absorbing all prefactors). Finally, $K$ is the anisotropy contant that enforces the easy-plane anisotropy.  Minimization of this energy yields a magnetization direction ${\bm n} = - \hat z$, with $\hat z$ the unit vector in the $z$-direction, for fields $B>K$. In this regime the magnons disperse quadratically and have a gap $\sim B+K$. For $B<K$ the magnetization direction deviates from the $z$-direction and acquires a component in the $x-y$-plane. In that case we have that $n_z = 1-B/K$ with the $x-y$-component determined by normalization. This latter so-called polar phase can be interpreted a Bose-Einstein-condensed phase of magnons \cite{flebus2016} and will turn out to have linearly-dispersing magnons. 

The Landau-Lifschitz-Gilbert equation is rewritten as a dissipative Gross-Pitaevskii equation by introducting the complex field $\psi ({\bm x},t)$ by means of ${\bm n} = (\sqrt{\mu_B/2M_s} {\rm Re} [\psi]), \sqrt{\mu_B/2M_s} {\rm Im} [\psi], \mu_B|\psi|^2/M_s)$ which corresponds to a classical linearized version of the usual  Holstein-Primakoff transformation. We find that 
\begin{eqnarray}
\label{eq:gpe}
  i \hbar \left( \frac{\partial }{\partial t} + {\bm
v}_{\rm s} \cdot \nabla  \right) \psi &=& \left( -J_s \nabla^2 - \mu + g |\psi|^2 \right) \psi + \nonumber \\
&&+ \hbar \alpha \left( \frac{\partial }{\partial t}+ \frac{\beta}{\alpha}{\bm
v}_{\rm s} \cdot \nabla  \right) \psi~,
\end{eqnarray}
with the chemical potential $\mu = K-B$ and contact interaction $g=K \mu_B/M_s$. In the polar phase when $B<K$ and thus $\mu>0$ we insert $\psi = \sqrt{n_c} + \delta \psi$, with $n_c = \mu/g$ the effective condensate density, into the above. Linearizing with respect to $\delta \psi$ and $\delta \psi^*$ leads to two coupled equations for $\delta \psi$ and $\delta \psi^*$. Using the Bogoliubov ansatz we write $\delta \psi = u ({\bm x}) e^{-i \omega t}  - v^* ({\bm x}) e^{+i \omega t}$ from which we find
\begin{widetext}
\begin{equation}
\label{eq:bogoham}
 \left( \begin{array}{cc}
      \hbar \omega + i \hbar {\bm v}_s \cdot \nabla + J_s \nabla^2 -\mu + i \alpha \hbar \omega - \beta \hbar {\bm v}_s \cdot \nabla &-\mu\\
      -\mu&   - \hbar \omega - i \hbar {\bm v}_s \cdot \nabla + J_s \nabla^2 -\mu + i \alpha \hbar \omega - \beta \hbar {\bm v}_s \cdot \nabla
    \end{array} \right)  \left( \begin{array}{c}
      u \\
      v
    \end{array} \right) =0~.
\end{equation}
\end{widetext}
The above equations are, up to the dissipative corrections proportional to $\alpha$ and $\beta$, equivalent to the equation describing Bogoliubov excitations on top of a Bose-Einstein condensate flowing with velocity ${\bm v}_s$ \cite{garay2000,book}. Note that here the velocity is not given by a superflowing condensate but by the electrons providing the nonzero charge current and resulting nonzero spin current. Taking, for the moment, ${\bm v}_s$ constant we find in the long-wave limit and to leading order in $\alpha$ and $\beta$ the magnon dispersion relation 
$ (\omega_{\bm k} - {\bm v}_s \cdot {\bm k }) =  c k - i \alpha  c k -  i (\alpha - \beta)  {\bm v}_s \cdot {\bm k}$, with the spin-wave velocity $c = \sqrt{2 J_s (K-B)}/\hbar $.

\begin{figure}[!t]
\begin{center}
\includegraphics[width=1.00\linewidth]{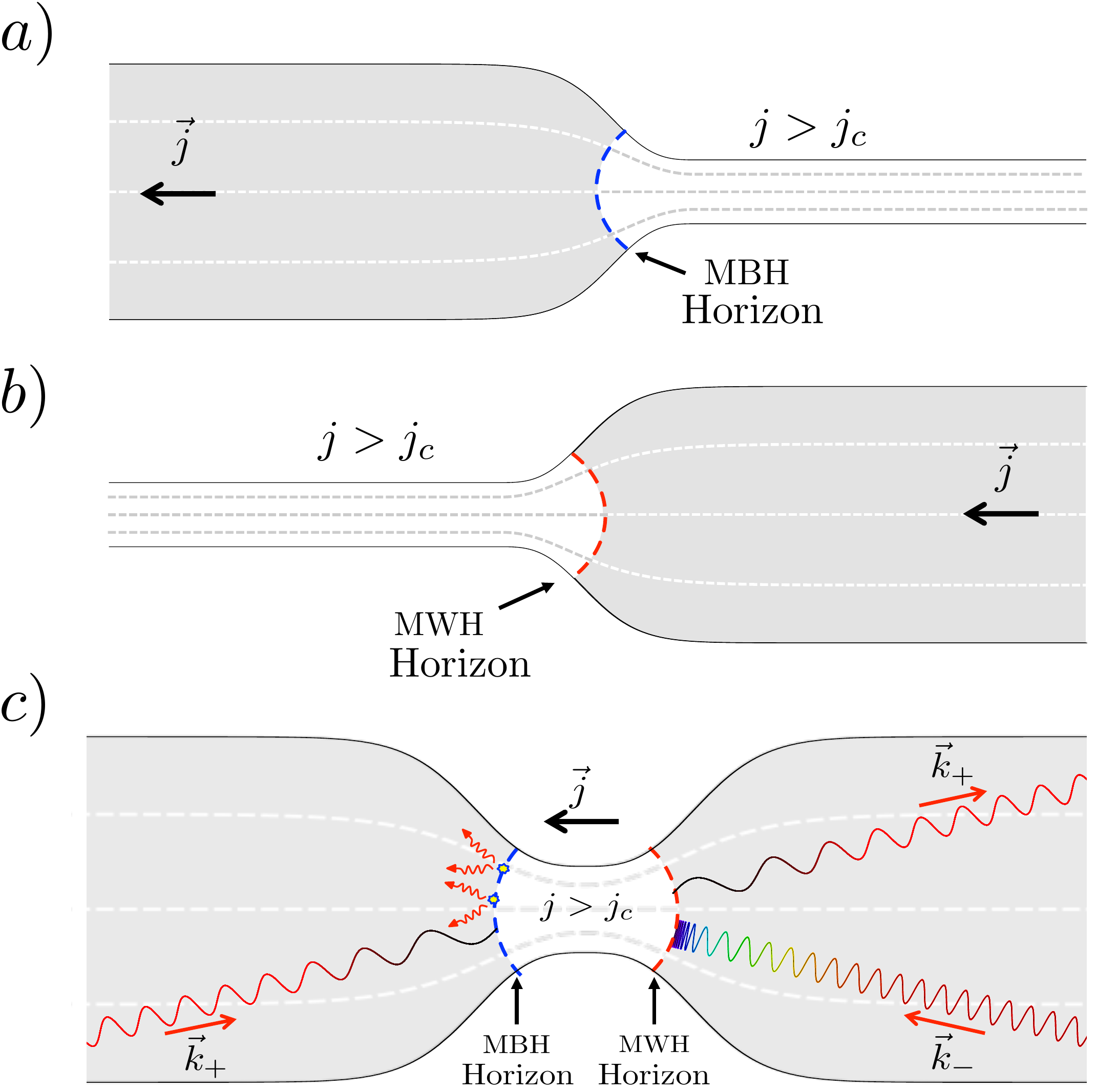}
\caption{(Color online) Set-up for creating magnonic black-hole and white-hole horizons. A narrow region of a wire that carries a steady-state current leads to an increased current density such that the background velocity $\propto - {\bm j}$ exceeds the magnon velocity if the current exceeds $j_c$. a) A magnonic black-hole (MBH) horizon for magnons incoming from the left. b) A magnon white-hole (MWH) horizon for magnons incoming from the right. c) A pair of MBH and MWH horizons. Incoming and scattered spin waves are illustrated.} \label{fig:scheme}\end{center}
\end{figure}

Black-hole and white-hole horizons are now implemented by regions where the velocity ${\bm v}_s$ changes from $|{\bm v}_s| <c$ to $|{\bm v}_s|>c$, and vice versa. We specifically consider the set-ups in Fig.~\ref{fig:scheme} that involve a wire geometry. A current-density is flowing from right to left, such that the velocity ${\bm v}_s$ is typically pointing from left to right (depending on the sign of the spin polarization $P$).  A narrow region in a wire leads to an increase in the current density and thus in $|{\bm v}_s|$ with respect to a wider region. If the current density in this narrow region is such that in the narrow region $|{\bm v}_s| >c$, while in the wider regions of the wire we have $|{\bm v}_s|<c$, there will be a black-hole horizon for magnons coming from the wide region on the left (travelling ``downstream", i.e., to the right) in Fig.~\ref{fig:scheme}~a). Similarly, there will be a white-hole horizon for magnons travelling to the left from the right ``upstream" region in Fig.~\ref{fig:scheme}~b). A dent in the wire creates a pair of horizons, a black and a white one [see Fig.~\ref{fig:scheme}~c)].

Since $\alpha$ and $\beta$ are small we ignore in first instance dissipation. We focus in the following on black-hole horizons. The Hawking temperature $T_H$ of the black-hole horizon is then given by  $k_B T_H = \frac{\hbar}{2\pi} \partial (|{\bm v}_s-c|)/\partial r $ \cite{book}, where the derivative is taken at the horizon and in the direction perpendicular to it, and where $k_B$ is Boltzmann's constant. Taking a typical value of $J_s = 10^{-39}$ J m$^{-2}$ for the exchange interactions, and $B/k_B$ and $K/k_B$ to be of the order of $1$ K \cite{parkin2015}, we estimate $c \sim 10^3$ m/s, although it can be made arbitrarily small by tuning $B \uparrow K$. The critical current density $j_c$ required for $|{\bm v}_s|$ to exceed $c$ is $j_c \sim M_s |e| c/\mu_B \sim 10^{11}$ A/m$^2$, where we took $g\sim P\sim 1$, $M_s/\mu_B \sim 1$ nm$^{-3}$, and $c=1000$ m/s. We note here that such large, or even larger, current densities are quite common in experiments on current-driven domain wall motion \cite{parkin2015}. Assuming now that the current density changes over a length scale of a $d=1$ nm - which can be achieved by nanofabrication techniques - we find that $T_H \sim \hbar c/k_B d \sim 1$ K.  
At zero temperature, pairs of magnons are created with one magnon being absorbed by the black hole. The black-hole horizon will emit magnons with a thermal spectrum determined by $T_H$ into the subsonic region left of the magnonic black-hole horizon in Fig.~\ref{fig:scheme}~a). Of course, the current density leads to an increase in temperature because of Joule heating such that zero or small temperatures are difficult to achieve. By tuning the field $B$ to approach $K$ one can lower $c$  and the required critical current $j_c$.  The Hawking temperature will go down accordingly, but the Joule heating is quadratic in temperature whereas the change in Hawking temperature is linear, allowing to disentangle both effects. 

There are also signatures of the physics of Hawking radiation in the classical regime, i.e., at temperatures $T \gg T_H$, as the underlying processes are linear. Following the arguments of Ref.~\cite{Weinfurther2011} we have that the ratio of  spin-wave  transmission ($t$) and reflection ($r$) amplitudes off the black-hole horizon is given by
\begin{equation}
\label{eq:reflandtrans}
  \frac{|t(\omega)|^2}{|r(\omega)|^2} = \exp \left(-\frac{\hbar \omega}{k_B T_H} \right)~.
\end{equation}
Spin-wave scattering experiments are standard in the field of magnonics \cite{kurglyak2010} and may thus provide a first step towards observing the non-trivial features of magnonic black-hole horizons. The presence of the horizon itself can of course also be detected with a spin-wave scattering experiment. 

Using the above expression for the transmission and reflection coefficients, and standard Landauer-B\"utikker expressions for magnon transport properties we find that magnon transport coefficients are proportional to $I_n =\int d \epsilon  \epsilon^n |t(\epsilon)|^2 \left(- \partial n_B/\partial \epsilon \right)$ with $n_B (\epsilon) =[e^{\epsilon/k_B T}-1]^{-1}$ the Bose-Einstein distribution function at the temperature $T$. Here, $I_0$ is proportional to the spin conductance and $I_1$ to the magnon contribution to the heat conductance. Using Eq.~(\ref{eq:reflandtrans}) and conservation of norm we conclude that, at low temperatures, the transport coefficients behave as if the actual temperature $T$ is replaced by $T^*$, with $1/T^*=1/T+1/T_H$, so that $T^*/T=(T_H/T)/(1+T_H/T)$. For the purpose of this estimate we have replaced the Bose-Einstein distribution function by the Boltzmann one. When $T_H \ll T$ the transport coefficients thus behave as if the temperature is equal to $T_H$, while in the opposite limit $T^* \approx T$. This may provide a transport signature of the Hawking radiation.  

When $|{\bm v}_s| > c$ the ferromagnetic ground state may become unstable towards the formation of a modulated state \cite{rossier2004, tatara2004}. This will not affect the physics in the subsonic region (i.e., left of the black-hole horizon in Fig.~\ref{fig:scheme}). Taking into account the dissipative terms we find that the magnons are linearly stable, however, when $|(\alpha-\beta) {\bm v}_s| < \alpha c$, which provides a large window for stability since $\alpha$ and $\beta$ are usually approximately equal.

{\it Isotropic antiferromagnetic metal} --- Our next proposal concerns an isotropic antiferromagnetic metal and may also be implemented using a synthetic antiferromagnet, i.e., two ferromagnetic layers separated by a normal metal, provided the interlayer exchange coupling is sufficiently strong. The interaction between spin-polarized current and the magnetization dynamics in antiferromagnets has been studied theoretically over the past decade \cite{nunez2006,swaving2011,hals2011}. Recently, electrical switching of an antiferromagnet was reported using strong spin-orbit coupling \cite{zelezny2014,wadley2016}. In the opposite limit of strong exchange interactions between electron spins and magnetization the equation of motion for the N\'eel vector ${\bm n}$ of the antiferromagnet is given by \cite{nunez2006,swaving2011,hals2011}
 \begin{equation}
 \label{eq:neel}
 {\bm{n}}\times(\ddot{{\bm{n}}}-c_a^{2}{\nabla^2{\bm{n}}}+(\bm{v}\cdot\nabla)^2{\bm{n}}+2(\bm{v}\cdot \nabla) \dot{{\bm{n}}})={\bm n} \times {\bm h}_d~,\label{eq: Eq. Motion Full}
 \end{equation}
where the velocity ${\bm v}$ plays the same role for antiferromagnets as the velocity ${\bm v}_s$ for ferromagnets. Furthermore, ${\bm h}_d \propto (\beta\ ({\bm v}\cdot \nabla) {{\bf{n}}}/\alpha+\dot{\bf{n}}) $ describes relaxation and will, like for ferromagnets, be ignored in the first instance since it is usually small.  Generally, we have that ${\bm v}  \propto {\bm j}$ with the prefactor determined by microscopic physics. Estimates \cite{swaving2011,Yamane2016} show that similar velocities as for case of ferromagnets can be obtained, i.e., $|{\bm v}|$ can be of the order of $1000-10000$ m/s for current densities ${\bm j} \sim 10^{11-12}$ A/m$^2$. Since the antiferromagnetic spin-wave velocity $c_a$ is of the order of $c_a \sim 1000$ m/s, we conclude that black-hole horizons can be created for magnons in antiferromagnets, as $|{\bm v}|$ can exceed $c_a$.

Linearizing Eq.~(\ref{eq:neel}) around a collinear state ${\bm n}_0$ by means of ${\bm n}={\bm n}_0+\delta {\bm n}$ we find that 
\begin{equation} \label{eq: Phi}
\frac{1}{c_a^2}\left(\frac{\partial}{\partial t}+\bm{v}\cdot\nabla\right)^2 {\bm{\Phi}}-\nabla^2{\bm{\Phi}}
=0~,
\end{equation}
where ${\bm \Phi}={\bm{n}}_{0}\times\delta{\bm{n}}$. This equation shows that antiferromagnetic magnons interacting with a transport current are described analogously to sound waves propagating in a medium with nonzero velocity, albeit it that antiferromagnetic magnons have two polarizations. Following the arguments of Ref.~\cite{Unruh} and considering the set-up in Fig.~\ref{fig:scheme} we find the same expression for the Hawking temperature as for the ferromagnetic metal (with $c$ replaced by $c_a$ and ${\bm v}_s$ replaced by ${\bm v}$). The possibilities for experimental detection in the classical and quantum regimes are also similar.

{\it Easy-plane magnetic insulators} --- We discuss for completeness briefly also a proposal based on easy-plane ferromagnetic insulators. Since they can be viewed as spin superfluids \cite{halperin1969, sonin2010,bunkov2013}, this particular realization is very much akin to analogue black holes based on flowing superfluids \cite{book}. It is distinct from the metallic cases discussed above as it does not benefit from the interaction between magnons and electrons in the bulk and the resulting ease of control over the flow velocities ${\bm v}_s$ and ${\bm v}$.  

Consider a ferromagnetic insulator described by the energy in Eq.~(\ref{eq:energy}). Its equation of motion for the magnetization direction is the same as Eq.~(\ref{eq:LLGwithSTTs}) but with ${\bm v}_s=0$. We consider now the situation that $B<K$ and that the static magnetic texture has a nonzero winding, i.e., $\psi ({\bm x},t) = \sqrt{n_c} e^{i {\bm v}_0\cdot {\bm x}/\hbar} + \delta \psi ({\bm x},t)$. Linearizing the equation of motion with respect to $\delta \psi$ in the same way as for the ferromagnetic metal one ultimately  arrives at Eq.~(\ref{eq:bogoham}) with $\beta=0$ (as there is no electric current) and ${\bm v}_s$ replaced by ${\bm v}_0$. Magnonic black and white holes are created by regions where ${\bm v}_0$ changes from smaller than $c$ to larger than $c$ (or vice versa). 

While the velocity ${\bm v}_0$ does not stem from an electron spin current flowing through the bulk of the system, it may be controlled by exploiting the flow of spin currents across the boundary between a normal metal and a magnetic insulator \cite{takei2014,takei2014b}. Spin superfluids have an upper and lower critical current that limits the range of ${\bm v}_0$. Finally, we mention that since an easy-plane antiferromagnetic insulator is also a spin superfluid \cite{halperin1969} a similar phenomenology holds. 

{\it Discussion, conclusion and outlook} --- In conclusion, we have shown that the interaction between spin current and magnetization dynamics can give rise to black-hole and white-hole horizons for magnons. For the metallic easy-plane ferromagnet we have discussed the effect of relaxation and how it stabilizes the homogeneous magnetic ground state. While an extensive investigation into the effects of dissipation in the quantum regime is beyond the scope of this paper we expect that it gives rise to  a characteristic length scale $\hbar c/\alpha k_B T$ over which the system needs to be quantum-coherent to observe spontaneous magnon pair creation. In our discussions, we have neglected the effects of unwanted anisotropies that give the magnons a gap. Such anisotropies can be neglected as long as the gap is smaller than $k_B T_H$. Given the experimental control over anisotropies by doping, sample shape, and material composition, we expect that this will not pose a severe limitation. 

One of the most interesting aspects of Hawking radiation is that the emitted particle pairs are entangled. For our case of magnetic systems the quantity
\begin{equation}
 \langle \hat S^-_{-{\bm k}} \hat S^{-}_{\bm k} \rangle \langle \hat S^+_{-{\bm k}} \hat S^+_{{\bm k}} \rangle-\langle \hat S^+_{\bm k} \hat S^{-}_{\bm k} \rangle \langle \hat S^+_{-{\bm k}} \hat S^-_{-{\bm k}} \rangle~,
\end{equation}
exceeds its classical value of zero if the emitted pairs are entangled \cite{busch2014}. Here, $\hat S^+_{\bm k}$ and $\hat S^-_{\bm k}$ are the usual spin raising and lowering operators at magnon momentum ${\bm k}$, and the magnons forming the pairs are emitted with momenta $+{\bm k}$ and $-{\bm k}$. Using the results of Ref.~\cite{steinhauer2015} we expect that the above correlation function can in principle be measured from a spin-spin correlation function, e.g., by neutron scattering. In future work we will investigate possible quantum-information devices exploiting the entanglement between the magnonic Hawking partners. Other interesting directions for future research include the interaction of ferromagnetic solitons, i.e., domain walls, with the horizons, inclusion of strong spin-orbit coupling, and developing a transport theory that treat the horizons beyond the estimates made here. 

{\it Acknowledgements} ---  RD would like to acknowledge a discussion with W.G. Unruh during his Kramers professorship at the Institute for Theoretical Physics in Utrecht that contributed to the ideas presented here. We also acknowledge helpful interactions with Erik van der Wurff, Henk Stoof, and Stefan Vandoren. RD  is member of the
D-ITP consortium, a program of the Netherlands Organisation
for Scientific Research (NWO) that is funded by
the Dutch Ministry of Education, Culture and Science
(OCW). This work is  in part funded by the Stichting voor Fundamenteel
Onderzoek der Materie (FOM). ASN would like to thank funding from grants Fondecyt 1150072. ASN also acknowledges support from Financiamiento Basal para Centros Cient\'ificos y Tecnol\'ogicos de Excelencia, under Project No. FB 0807 (Chile).

\end{document}